# Transition from Analysis to Software Design: A Review and New Perspective

Hamdi A. Al-Jamimi*, Moataz Ahmed

Information and Computer Science Department
King Fahd University of Petroleum and Minerals
Dhahran, Saudi Arabia
Email: {aljamimi, moataz}@kfupm.eu.sa

*Abstract*—Analysis and design phases are the most crucial part of the software development life-cycle. Reusing the artifacts of these early phases is very beneficial to improve the productivity and software quality. In this paper we analyze the literature on the automatic transformation of artifacts from the problem space (i.e., requirement analysis models) into artifacts in the solution space (i.e., architecture, design and implementation code). The goal is to assess the current state of the art with regard to the ability of automatically reusing previously developed software designs in synthesizing a new design for a given requirement. We surveyed various related areas such as model-driven development and model transformation techniques. Our analysis revealed that this topic has not been satisfactorily covered yet. Accordingly, we propose a framework consists of three stages to address uncovered limitations in current approaches.

I. INTRODUCTION

In reality, the software development process can be seen as a series of different phases. Each phase, in this process, produces new models by utilizing the models built during its preceding phases. Analysis and design represent the most crucial part of the software development life-cycle. In this context, the earlier artifacts represent the problem space (i.e., the problem to be solved). These artifacts include the software requirements specifications (SRS), conceptual models, and analysis class diagrams. On the other hand, the solution space (i.e., the solution to the problem) can be represented by the subsequent artifacts including the architectural documents, detailed design class and sequence diagrams. Particularly, the *analysis phase* is to related to understanding the given problem, while the *design phase* is related to the formation of a solution for the analyzed problem [1]. The design task is seen as a more complex than the analysis phase. That is because most of the decisions made at the design, in turn this stage has a considerable influence on its subsequent phases. Therefore, design phase requires more knowledge and experience from the developers.

Software reuse can be conducted at any stage of the software development process. Various levels of reuse can be conducted; analysis, design, code, and test level [2]. Software reuse is considered as a promising way to improve software development productivity and quality. Software developers realize that reuse of early life-cycle artifacts constructed at the beginning of the software development life-cycle has its own importance where it allows utilizing all the related late artifacts during the software development.

The goal of this work is to move from the analysis models (defined as the problem space) toward the software design (defined as the solution space) by reusing previously developed software. That is, based on the given requirements the existing requirement-design pairs from previous systems can be utilized to build the new system's design. Such that the resultant design holds certain preferred quality properties. An overview of intended process is demonstrated in Figure 1. The main idea of this work is a part of development of environments integrated with CASE tools and capable facilitating early-stage artifacts reuse [3]. A major focus of the Integrated Reuse Environment (IRE) is to offer tools to facilitate reusing design and later artifacts based on matching requirements. In other words, for new requirements the IRE should facilitate assessing the similarity between new requirements to the requirements of completed projects to provide closest match so that their design counterparts can be reused with minimal effort.

In this paper, we introduce the problem of reusing previously developed designs to come up with a new design which is suitable for the presented requirement. In our search for a suitable solution to the specified problem of transition from software analysis to software design utilizing the analysis-design pairs, we reviewed many approaches in the literature. We explored model-driven development (MDD) and model transformation approaches including rules-based, pattern-based and example-based techniques. As a final point, we present a framework that consists of three stages to mine the repository with the aim of reuse, refine, and synthesis existing designs to come up with a design satisfying the new requirement.

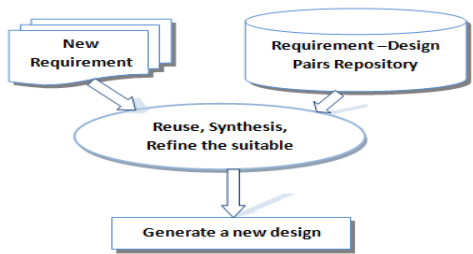

Fig. 1. High level view of proposed process

The rest of this paper is organized as follows; section II details the problem statement. A comprehensive literature survey for the possible solutions is introduced in Section III. Section IV describes the proposed framework to solve the described problem. Finally, Section V concludes the paper and presents the suggested future works.

## II. PROBLEM STATEMENT

The area of research mentioned earlier is surrounded with a number of difficulties.

First, in the existing widely used software development methodologies, it is noticeable the transition from the analysis artifacts to the design artifacts is unclear. Thus, obviously, there is a fundamental gap between the analysis phase and the design phase. The difficulty in moving from analysis to design is caused by the fact that the artifacts of analysis phase and design phase represent different things. While the analysis phase is linked to a human activity, the design phase is related more to the information technology systems [4]. For instance, although object-oriented paradigm is the dominant software development approach in the industry, it is still partially unclear how analysis relates to design in this paradigm. Even though, several benefits attributed to this paradigm, it has been unsuccessful in finding a way to systematically transition from the analysis phase to the design phase [4].

Second, the existing methodologies are general guidelines, thus software designers heavily rely on their experience from the development of previous systems to design new ones. As results, an essential part of the achieved transition relies more on the designer's experience and their subjective measures. In like this situations, they rely little on defined processes and methodologies.

The design phase is a multifaceted problem, where the design phase can be divided into two sub phases: conceptual design and detailed design. Since the design phase represents the solution space, so in the conceptual design the solution is analyzed in order to define the entities and subsystems that comprise the design model. This can be viewed as a high-level task where the main concerns are related with the specification phase and not with the implementation phase. Roughly speaking, to be able to satisfy the specification produced in the analysis phase, the conceptual design identifies the software architecture. On the other hand, the objective of the detailed design is to prepare the software's implementation phase. In turn, the detailed design model serves as a high-level view of the source code. The algorithms and the data structures are defined, as well as the organization and key features are described for the implementation phase.

What is noticeable is that, to translate the requirements into a high-level design the designers and developers spend significant amounts of time to accomplish this task. Although of that, there are methodologies that describe and manage requirements and design artifacts.

Last, reusing the previously developed designs is not trivial tasks especially when there is a possibility of selecting the appropriate building blocks from a variety of available designs and synthesis them to generate the intended design.

Indeed, this problem has two dimensions of difficulty; finding the needed blocks among all the blocks in the design, then combining these obtained blacks to represent a complete design.

Therefore, there is a real problem may encounter the developer in the three cases: first case, when adapting parts of design if none of the existing parts satisfies the need of reusing. Second case, when combine fragments of designs to come up with a design satisfying the given requirement. Third case, if none of the existing designs or parts of them satisfies the new requirements, there is no a clear mapping that guides the transition from the problem space (software analysis) to the solution space (software design).

Based on that, there is a need for intelligent tools that support the transition from software analysis to software design utilizing the analysis-design pairs. The purpose of these tools is not only to implement the common software reuse techniques, also to provide support for more complex reasoning abilities and exploration of new design spaces. Moreover, based on the observations from the previously developed systems, the intelligent tools may boost more creative designs. Therefore, the idea of this work is to building a framework that would provide assistance to the software designer, in tasks such as exploration of the design space.

As shown in Figure 1, the objective is to reuse, refine and generate a design for the new requirement by utilizing the available requirement-design pairs in the repository. This might be achieved by retrieving the corresponding design to the matched requirements, or by generate new design based on rules extracted for the existing examples. To extract such rules from the existing examples, "learning by examples" should be utilized. Moreover, in order to mine the repository for the suitable blocks or fragments of designs and to combine these collected fragments from different designs; machine learning techniques are needed for selection, permutation, and integration.

## III. LITERARTURE SURVEY

In this section we present a comprehensive literature survey where we surveyed different areas trying to find out a solution for the specified problem. The survey addresses two views: transition from analysis to design, and model driven development including the model transformations.

### A. Transition from Analysis to Design

In reality it is difficult to move from software analysis to software design automatically. Thus, recognizing the differences between what is modeled in the phases can help significantly to come up with a more conscious development approach.

Kaindl [4] studied analysis and design models of real-world projects, to validate his view about the difficulty of the transition from analysis to design of software. He emphasizes that, the transition between analysis (requirements phase) and software design is an issue regardless of whether developers use an object-oriented approach or not.

Analyzing the requirements and building the models of analysis and design are cumbersome and complicated tasks which require automated support. The Natural Language (NL) is used frequently to describe the software requirements. In a typical software industry, SRS is written in NL to enhance communication between different stakeholders. Due to its inherent ambiguity, it is particularly not easy task to generate design objects from NL specification. However, the structured and constrained NL can be utilized to improve the correctness of the design.

Most of the work related to the moving from requirements to analysis and then to design only focused on the first transition based on NL processing [5-17]. Some other studies tried to obtain class diagram form use cases [18-20], however the resultant class diagrams still in the high level description. Similarly, other researchers [21-23] tried to generate other analysis diagrams from use cases.

As stated by [24] most often software architecture is identified formally, while software requirements are captured informally. Therefore, there is a gap when transition from the requirements to architecture. In this regard, a substantial amount of research has been conducted to bridge this semantic gap. Grunbacher et al. [24] utilize intermediate models that are closer to software architecture to introduce mapping from requirements to architecture. For this purpose, they propose an approach called CBSP (abbreviation of Component, Bus, System, and Property). They have applied CBSP within the context of different requirements and architecture definition techniques. Liu et al. [25] analyze the gap between the software requirements and the software architecture to identify the inadequacy of mapping approaches in traditional structured method and object-oriented method. Based on that, they propose a feature-oriented mapping and transformation approach from requirements to software architecture. Kaindl et al. [26] suggest the use of model driven approaches to ease the mapping from the software requirements to the architectural design.

Despite of the scientific contributions of the mentioned studies, there is still lack of effective solutions. As shown in earlier work by Kaindl, object-oriented domain models cannot be simply become object-oriented design models. Neither is it possible to transform domain models to design models. There would be the implicit assumption that each and every object class in the domain model would finally end up as several object classes in the detailed design and consequently the implementation.

Larmen also states in [27] that domain models represent real-world concepts and not software objects and thus cannot be transformed automatically to a software design, but having mappings between domain and design classes lowers the representational gap between our mental model and the software. Even though automatic transformation seems not possible without intelligent problem solvers that establish traces (like in [28]) and mappings between a domain and design model. The mappings from requirements to design may be viewed as special and elaborate forms of traceability links.

All of these concerns motivated Kaindl and Falb [29] to ask "Can the transition from requirements to software design be a model-driven transformation or just a mapping?".

Based on that, they further discussed whether model-driven transformations are appropriate and applicable for moving from requirements to software design.

*B. Model Driven Development*

MDD is an emerging software development technology introduced for the purpose of bridging the gap between the problem space and solution space. In this technology, the models are considered the primary artifacts of the development process; also they contain the needed information that supports its different phases. This sequence of models can be created, refined and maintained. Hence, software designers and developers can concentrate on high-level problem solving rather than focusing on low-level implementation details [30].

MDD supports the reuse through different levels of abstraction provided by the models at different stages of the development life cycle. It distinguishes between three types of models:

- Computation Independent Model (CIM) focuses on the domain a higher level of abstraction instead of showing the details of the system structure.
- Platform Independent Model (PIM) designed without considering the underlying platform or any other technical considerations.
- Platform Specific Model (PSM) includes the technical considerations and the underlying platform.

**CIM to PIM**- The transformation from CIM to PIM has no a lot of attention of the researchers, as well as there is no comprehensive literature survey available in the specific domain of CIM to PIM transformation. Reviewing and analyzing the existing approaches would facilitate determining the gaps, weaknesses and needed enhancement for this kind of transformation. In addition, it might give an idea about the automatic moving from CIM to PIM if possible. Table 1 summarizes a review and comparison for the existing CIM-to-PIM transformation approaches based on different evaluation criteria which include the following:

1. CIM consists of two aspects: the business process model (BP) shows all the business activities, and the requirement model (RM) which specifies the system.

2. CIM representation: UML, BP notations (BPN), and Data Flow Diagram (DFD) are used to represent the business process. Use Case (UC) and Feature model (FM) are used to describe the requirements.

3. PIM aspects: include functional (F), structural (S), and behavioral (B) perspectives.

4. PIM representation: UC, Activity Diagram (AD), Sequence Diagram (SD), and Class Diagram (CD).

5. Transformation Mechanism used for transition.

6. Automation: to which extent the proposed approach was automated: fully or partially automated.

TABLE I.    SUMMARY OF THE CIM-TO-PIM TRANSFORMATIONS

| Study | Evaluation Criteria | | | | | |
|---|---|---|---|---|---|---|
|  | *CIM aspects* | *CIM Representation* | *PIM aspects* | *PIM Representation* | *Transformation Mechanism* | *Automated* |
| Zhang et al. [31] | RM | FM | S | SW Arch | Responsibilities | Semi |
| Kardoš et al. [32] | BM | DFD | F,B,S | UC, SD, AD, CD | Manual | Manual |
| Kerraf et al. [33] | BM, RM | AD | S | CD | Manual | Manual |
| Cao et al. [34] | RM | FM | S | SW Arch | Patterns | Semi |
| Castro et al. [35] | BM, RM | e3 value model BPMN | F,B | UC, Service process | Meta-models Mappings –ATL | Partial |
| Rodríguez et al. [36] | BM | BPMN | F | UC | QVT and refinements rules, checklists | Semi |
| Raj et al. [37] | BM | SBVR | B,S | AD, SD, CD | Manual | Manual |
| Suarez et al. [38] | BM | AD | S | CD | Manual | Manual |

**PIM to PSM** – Since PIM reflects the features of the problem domain, the model is transformed into PSM in order to implement the PIM. That is, to consider implementation issues and the underlying platform [30]. PSM may contain features that are presented in PIM, thus PSM is not necessarily a refinement of the PIM [39]. The platform-specific details need to be generated using different tools in order to automate the generation of those details. For defining transformations, those tools offer three different approaches: first, direct model manipulation which can operate on a set of procedural APIs. Second, intermediate representation deals with models in a standard form such as XML. Third, transformation language support which expresses, composes, and applies transformation explicitly [40].

In this regard, many PIM to PSM transformations studies have been conducted in the literature. The model-to-model transformation approaches can be categorized into: operational and declarative. The first category is based on rules that explicitly identify how to create the target models elements starting from the source model elements. The second category gives a explanation of the mappings between the source and target models focusing on the relation hold between two models.

Informative surveys of model transformation languages can be found in [41-44]. Due to the limitation of the paper size we just refer to the previous surveys that have been conducted on this topic [41-43]. Czarnecki et al. [41] classify hierarchically the specification of model transformation approaches based on feature diagrams into a number of classes. The feature model offers a terminology used to describe the model transformation approaches as well as to make the different design choices for such approaches explicit. Mens et al. [42] provide a multi-dimensional taxonomy of model transformations. The introduced taxonomy is more targeted towards techniques and formalisms supporting the activity of model transformation. The main purpose of this taxonomy is to position model transformation tools and techniques within the domain; as well as to identify and evaluate tools or technologies for a specific model transformation activity.

A conclusion to be drawn from studying the existing rule-based and pattern-based transformation approaches is that they are often based on empirically obtained rules. When identifying the transformation rules and automating the transformation process most of the researcher follow a common approach which is the use of a model transformation language. These languages still suffer some limitations, although most of these languages are able to implement complex and large-scale model transformation tasks. The users may encounter some challenges when dealing with specific transformation language specially the users who are unfamiliar with that language.

In addition, it might be a difficult task to define, express, and maintain the transformation rules, particularly for non-widely used formalisms. Another dimension of difficulty may appear when the declarative expressions are not at the proper level of abstraction for an end-user. This may affect negatively the learning curve and training cost.

Moreover, since the transformation rules are usually defined at the meta-model level, there is a need to understand well the abstract syntax of the source and target models. The semantic interrelationships between these models also need to be known. However, in some situations, it is difficult to expose the domain concepts because they might be hidden in the meta-model.

These implicit concepts make writing transformation rules demanding. Accordingly, some domain experts may encounter difficulties when trying building model transformations for the domain in which they have extensive experience. This because of the difficult when specifying transformation rules at the meta-model level, and the associated learning curve.

*C. Example-Based Model Transformation*

To tackle the mentioned negative aspects of the rule-based and pattern-based model transformations, a number of example-based approaches have been proposed for model transformations. Example-based model transformation (EBMT) is a recent trend of research aiming at learning a transformation between the source and target from existing examples. The form of the transformation example is specified by a source model, a target model and mappings between source elements and the corresponding target elements. EBMT allows defining transformations using examples represented in concrete syntax instead of using the computer internal representation of models.

In general, here we can distinguish between two kinds of approaches. First, the demonstration-based approaches where the model transformation is demonstrated in the modeling editor. The example models are modified. Then the resultant modifications are recorded. The general transformation is derived from the concrete changes, then it may be replayed on other models. Second, the example-based approaches where the input, output models, and the correspondences between them are given by the user rather than demonstrating the transformation in modeling editors.

Several Model Transformation By-Demonstration (MTBD) approaches have been proposed for reducing the effort of writing model transformation rules manually. MTBD approaches record actions performed on example models to derive general operations. Approaches of MTBD only for in-place transformations proposed by Sun et al. [45] and Brosch et al. [46].

MDD aims to use platform independent modeling techniques in order to abstract from the implementation level of software systems. On the other hand, the aim of by-example approaches is simplicity the development of systems. Instead of the direct developing, it is possible to utilize the existing examples to draw a clear map. Again, in MDD different transformation scenarios occur between the various models, thus different by-example approaches can be employed for these transformations. Therefore, it is worthy and promising idea to merge both paradigms.

Recently, a number of EBMT approaches have been proposed such as [47-51]. Kappel et al. [52] introduce an overview about the different example-based approaches. They divide them into two categories: demonstration-based and correspondence-based approaches. For each, they discuss their concepts and previously proposed techniques.

## IV. THE PROPOSED FRAMEWORK

This section details the proposed framework addressing the aforementioned limitations in the current approaches. In order to generate the appropriate design for the given new requirement, three different stages have been proposed as Figure 2 depicts in details.

For the given problem "new requirement" it is supposed to obtain a solution "appropriate design" in one of the three different stages. Means that when obtaining a solution in the first stage, the process terminates and return the acquired solution, without need for trying the second and third stages. Similarly when a solution is obtained in the second stage, saves us from moving to the third stage.

Assuming we have previously developed requirement-design pairs repository that can be utilized to build a rule base. Machine learning help extract the desired knowledge from existing examples and extensively ease the development of formal rules. Thus, some machine learning techniques can be utilized to find mappings between the requirements and design pairs, based on the available examples. The mappings can be recognized and formalized as rules stored in the rule base to be applied later in the third stage in our proposed framework.

*A. First stage*

Retrieval process can be viewed as two tasks: retrieval of similar requirements from the requirements-design pairs (the retrieval task), and ranking of these requirements against the given problem (the ranking task). The retrieval task is usually performed based on matching process using similarity metrics. Each of the retrieved requirement-design pair has a degree of similarity to the given requirement. If, the retrieved requirement from the repository shows a satisfied matching which is above a predefined threshold, then the corresponding design can be retrieved to be the new design for the given problem. Otherwise, in case the retrieved pairs do not show a high matching, then we move to the second stage.

*B. Second stage*

If the similarity metrics shows unsatisfied matching to the requirements stored in the repository then the new requirements will be compared against the generic analysis model.

Indeed, the generic analysis model contains all the patterns as well as the variations of the stored analysis models. Thus, we need to perform permutations by utilizing one of the heuristic search techniques (such as genetic algorithms, simulated analyzing or particle swarm algorithm) to produce an instance of the generic analysis model that excludes or includes some of the variation. Afterwards, the produced instance of the generic model is evaluated against the given requirement. The matching result is evaluated again, and the permutation is produced again and evaluated till we get the desired similarity percentage. Then the mapping from the instance, showing the best matching, to the generic design model can be utilized to retrieve the corresponding design from the generic design model. It supposed that, the generic design model contains all the patterns and the variations of the stored design models.

In case the instances of the generic analysis model didn't show satisfied matching to the given requirements, then the third stage is applied in order to combine and refine various designs existing in the repository.

*C. Third Stage*

This stage consists of three main parts:

First part, the ranked pairs retrieved form stage one are used to generate the design. Matching and merging is applied on the top of the ranked pairs to produce an initial design. Due to the huge search space and the enormous number of possibilities that can be utilized for merging, definitely heuristics algorithms are needed for this purpose.

Second part of this stage, relies on applying the rules, extracted from the examples, to improve the initial design. The outcome of this part is improved design that is presented to an expert to evaluate the applied rule.

Third part of this stage is the validation of the rules to update the rules and provide feedback to the rule base. Reinforcement learning can be utilized in this part. A reward function, based on the consequences of applying the rules, will evaluate the rule by numeric rewards and punishments in order to maximize its rewards and to learn next time to which extent the rule can be applied or ignored.

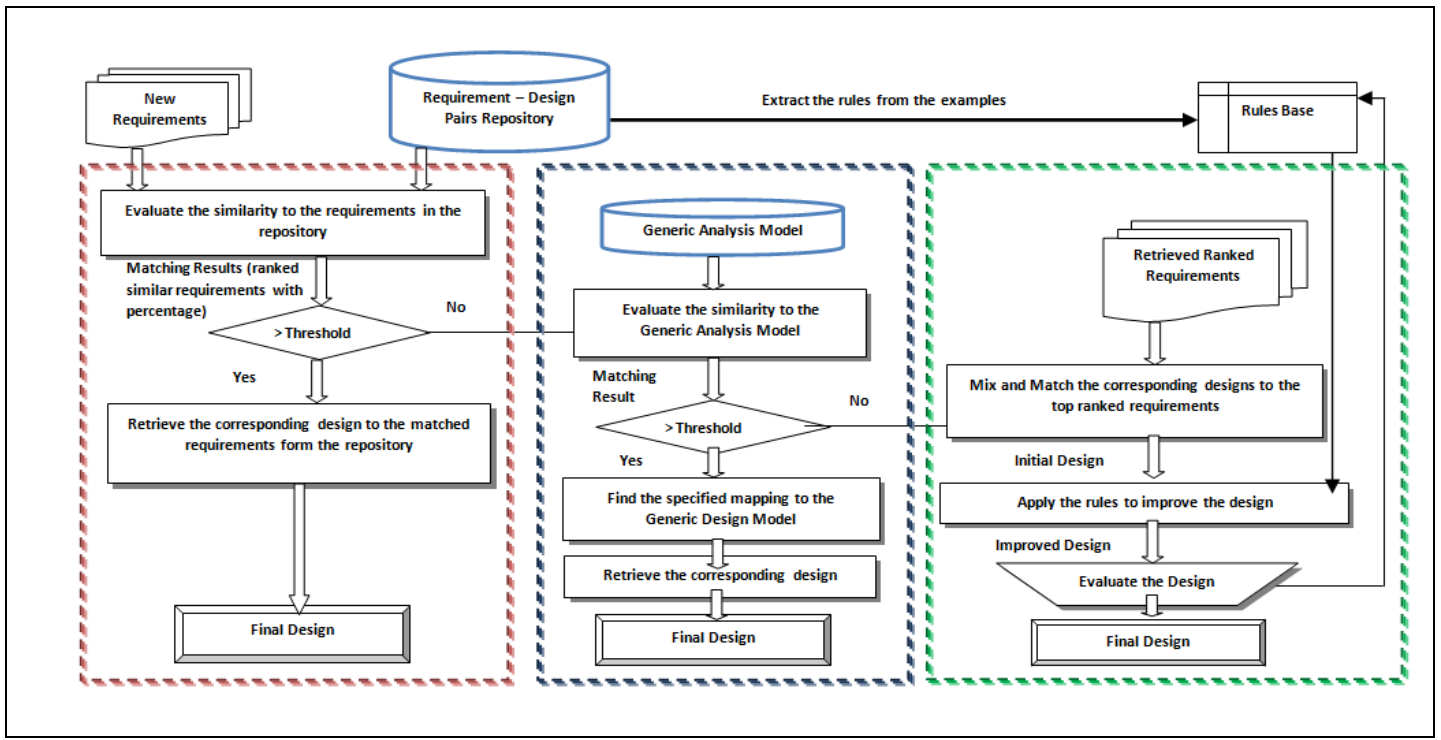

Fig. 2.    An overview of the proposed framework.

## V. CONCLUSION

In this paper, we introduced the problem of reusing previously developed designs to come up with a new design which is suitable a solution for a given presented requirement. We explored several research areas that may bring in a satisfied solution. The areas facilitate the transition from analysis to design such as model-driven development, model transformation including rules-based, pattern-based and example-based techniques. Accordingly, we proposed a framework that consists of three stages to mine the repository with the aim of reuse, synthesis, and refine the exits design to come up with a design satisfying the new requirement. Machine learning techniques and reinforcement learning are needed to accomplish the proposed solution and be able to synthesis and refine a design for a given requirement.

### ACKNOWLEDGMENT

The authors would like to acknowledge the support provided by the Deanship of Scientific Research at King Fahd University of Petroleum and Minerals, Saudi Arabia, under Research Grant 11-INF1633-04.

\* Corresponding Author:
Hamdi Ali Al-Jamimi,
Information and Computer Science Department,
King Fahd University of Petroleum and Minerals,
Dhahran, Saudi Arabia.
 Email: aljamimi@kfupm.edu.sa Tel: +966-596270853